\documentclass[letterpaper,twocolumn,10pt]{article}
\usepackage{graphics}
\usepackage{usenix}

\usepackage{url}
\usepackage{verbatim}
\usepackage{amssymb}
\newcommand{\msg}[1]{{\small \textsf{#1}}}

\newcommand{\eat}[1]{} 

\begin{document}

\date{}
\title{\Large \bf Attrition Defenses for a Peer-to-Peer Digital Preservation System}

\author{
TJ Giuli\\
\small{Stanford University, {CA}}
\and
Petros Maniatis\\
\small{Intel Research, Berkeley, {CA}}
\and
Mary Baker\\
\small{HP Labs, Palo Alto, {CA}}
\and
David S. H. Rosenthal\\
\small{Stanford University Libraries, {CA}}
\and
Mema Roussopoulos\\
\small{Harvard University, Cambridge, {MA}}
}

\maketitle

\thispagestyle{empty}

\begin{abstract}
In peer-to-peer systems, \emph{attrition attacks}
include both traditional, network-level denial of service attacks as
well as application-level attacks in which \emph{malign} peers conspire to waste
\emph{loyal} peers' resources.
We describe several defenses for LOCKSS, a peer-to-peer digital
preservation system, that help ensure that application-level attacks even
from powerful
adversaries are less effective than simple network-level attacks, and that
network-level attacks must be intense, wide-spread, and prolonged to
impair the system. {\tiny\verb!$Revision: 1.317 $!}
\end{abstract}

\section{Introduction}
\label{sec:introduction}

Denial of Service (DoS) attacks are among the most difficult for
distributed systems to resist.
Distinguishing legitimate requests for service from the attacker's requests
can be tricky,
and devoting substantial effort to doing so can easily be self-defeating.
The term DoS was introduced by Needham~\cite{Needham1993short} with
a broad meaning but over time it has come to mean high-bit-rate
network-level flooding attacks~\cite{Hussain2003short} that
rapidly degrade the usefulness of the victim system.
In addition to DoS, we use the term \emph{attrition}
to include also moderate- or low-bit-rate
application-level attacks that gradually impair the victim system.

The mechanisms described in this paper are aimed at
equipping the LOCKSS\footnote{LOCKSS is a trademark of Stanford University. It stands for
``Lots Of Copies Keep Stuff Safe.''} peer-to-peer (P2P) digital preservation
system to resist attrition attacks.  The system is in use at
about 80 libraries worldwide; publishers of about 2000 titles
have endorsed its use. Cooperation among peers reduces the cost
and increases the reliability of preservation, eliminates the
need for backup, and greatly reduces other operator interventions.

A \emph{loyal} (non-malign) peer participates in the LOCKSS system for two reasons:
to achieve regular reassurance that its content
agrees with the consensus of the peers holding copies of the same content,
and if it does not, to obtain the needed repair.
The goal of an attrition adversary is to prevent
loyal peers from successfully determining the consensus of their
peers or from obtaining requested repairs for so
long that undetected storage problems such as
natural ``bit rot'' or human error corrupt their
content. Other types of resource waste may be inconvenient
but have no lasting effect on this system.

We have developed a set of defenses,
some adapted from other systems, whose
combination in a P2P context provides novel and effective protection against
attrition.  These defenses
include \emph{admission control}, \emph{desynchronization}, 
and \emph{redundancy}.
Admission control, effected via rate limitation, first-hand
reputation, and effort balancing, ensures that legitimate requests can
be serviced even during malicious request floods.
Desynchronization ensures that progress continues even if
some suppliers of a needed service are currently too busy.
Redundancy ensures that the
attacker cannot incapacitate the system by targeting only few peers
at a time.

In prior work~\cite{Maniatis2003lockssSOSPshort} we
used redundancy and rate limitation to defend
LOCKSS peers against
attacks seeking to corrupt their content.
That system, however, remained vulnerable
to application-level attrition;
about 50 malign peers could abuse the protocol to prevent a network of
1000 peers from auditing
and repairing its content.

This paper presents a new design of the LOCKSS protocol that makes four contributions.
First, we demonstrate how our new design ensures that
application-level attrition, no matter
how powerful the attacker, is less effective than simple network
flooding. We do this while retaining our previous resistance against
other adversaries.  Second, we show that
even network-level attacks that continuously prevent \emph{all}
communication among a majority of the peers must last for months
to affect the system significantly.  Such attacks are many orders of
magnitude more powerful than those
observed in practice~\cite{Moore2001short}.
Third, since resource management lies at the
crux of attrition attacks and their defenses,
we extend our prior evaluation~\cite{Maniatis2003lockssSOSPshort}  to
deal with numerous concurrently preserved archival units of content
competing with each other for resources.
Finally, resource over-provisioning is essential in defending against
attrition attacks.
Our contribution is the ability to put an upper bound
on the amount of over-provisioning required to
defend the LOCKSS system from an arbitrarily powerful attrition adversary.
Our defenses may not all be immediately applicable
to all P2P applications, but we believe that many systems may benefit
from a subset of defenses, and that our analysis of the
effectiveness of these defenses is more broadly useful.

In the rest of this paper,  we first describe our application. We
continue by outlining how we would like this application to behave
under different levels of attrition attack.  We give an overview of the LOCKSS
protocol, describing how it incorporates each of our attrition defenses.
We then
explain the results of a systematic exploration of simulated attacks
against the resulting design,
showing that it successfully defends against attrition attacks
at all layers, from the network level up through the application protocol.

\section{The Application}
\label{sec:protocol}

In this section, we provide an overview of the digital preservation
problem for academic publishing, the problem that LOCKSS seeks to
solve. We then present and justify the set of design goals required of
any solution to this problem, setting the stage for our approach in
subsequent sections.

Academic publishing has migrated to the Web~\cite{Tenopir2004short},
placing society's scientific and cultural heritage at a variety of
risks such as confused provenance,
accidental editing by the publisher,
storage corruption, failed backups,
government or corporate censorship,
vandalism, and deliberate rewriting of history.
The LOCKSS system was designed~\cite{Rosenthal2000short} to
provide librarians with the tools they need to preserve
their community's access to journals and other Web materials.

Any solution must meet six stringent
requirements.  First, since under US law~\cite{EFF-DMCAshort} copyright Web content can only
be preserved with the owner's permission, the solution must accommodate
the publishers' interests.  Requiring publishers, for example, to offer perpetual no-fee access or
digital signatures on content makes them reluctant to give that permission.
Second, a solution must be extremely cheap in terms of
hardware, operating cost, and human expertise.
Few libraries could afford~\cite{ARLstats} a solution involving handling
and securely storing off-line media, but most can afford the few cheap
off-the-shelf PCs that provide sufficient storage for tens of thousands of
journal-years.  Third, the existence of cheap, reliable storage cannot
be assumed; affordable storage is unreliable~\cite{PCActive2003,Rosenthal2004short}.
Fourth, a solution must have a long time horizon.
Auditing content against stored digital signatures,
for example, assumes not only that the cryptosystem will remain unbroken,
but also that the secrecy, integrity, and availability of the keys are guaranteed 
for decades.  Fifth,
a solution must anticipate adversaries capable of powerful attacks
sustained over long periods; it must withstand these attacks, or at
least degrade slowly and gracefully while providing unambiguous
warnings~\cite{Rodrigues2004short}.  Sixth, a solution must not require a central locus of control
or administration, if it is to withstand concentrated
technical or legal attacks.

Two different architectures have been proposed for preserving
Web journals.
On one hand, trusted third party archives require publishers to
grant the archive permission, under certain circumstances, to republish
their content.  It has proved very difficult to persuade
publishers to do so~\cite{Cantara2003short}.
In the LOCKSS system,
on the other hand,
publishers need only grant their subscribing libraries permission to
supply their own content replica to their local readers.
This has been the key to obtaining permission from publishers.
It is thus important to note that our goal is not to minimize
the number of replicas consistent with content safety.
Instead, we strive
to minimize the per-replica cost of maintaining a large number
of replicas.  We trade extra replicas for fewer
lawyers, an easy decision given their relative costs.

The LOCKSS design is extremely conservative, making few
assumptions about the infrastructure. Although we believe this is
appropriate for a digital preservation system, less conservative assumptions
are certainly possible.  Taking increased risk
can increase the amount of content that can be preserved with given
computational power.  For example, the availability of limited amounts
of reliable, write-once memory would allow audits against local
hashes, the availability of a reliable public key infrastructure might
allow publishers to sign their content and peers to audit against the
signatures, and so on.  Conservatively, the assumptions underlying such
optimizations could be violated without warning at any time;
the write-once memory might be corrupted or mishandled
or a private key might leak.
Thus,
designs using these optimizations would still need the audit
mechanism as a fall-back.
The more a peer operator can do to avoid local failures
the better the system works,
but our conservative design principles lead us to focus on mechanisms
that minimize dependence on these efforts.

With this specific application in mind, we tackle the ``abstract'' problem of
auditing and repairing replicas of distinct \emph{archival units} or AUs
(a year's run of an on-line journal, in our target application) preserved by a population of peers
(libraries) in
the face of attrition attacks.  For each AU it preserves, a peer starts out with its own,
correct replica
(obtained from the publisher's Web site), which it can only use
to satisfy local read requests (from local patrons) and to
assist other peers with replica repairs.  In the rest of this
paper we refer to AUs, peers, and replicas, rather than journals and libraries.

\section{System Model}
\label{sec:attritionAnalysis}

In this section we present the adversary we model, our security goals
for the system, and our defensive framework.

\subsection{Adversary Model}
\label{sec:attritionAnalysis:adversaryModel}

In keeping with our conservative design philosophy, we assume a powerful
adversary with several important abilities.
\emph{Pipe stoppage} is his ability to prevent communication with
victim peers for extended periods by flooding links with garbage packets
or using more sophisticated
techniques~\cite{Kuzmanovic2003short}. \emph{Total information
awareness} allows him to control and monitor all of
his resources instantaneously. He has \emph{unconstrained identities} in that
he can purchase or spoof unlimited network identities.
\emph{Insider information} allows him
complete knowledge of his victims' system parameters and
resource commitments. \emph{Masquerading} means that loyal peers cannot
detect him, as long as he follows the 
protocol. Finally, he has \emph{unlimited computational resources},
though he is polynomially bounded in his computations (i.e., he cannot invert
cryptographic functions).

The adversary employs these capabilities in \emph{effortless} and
\emph{effortful} attacks.  An effortless attack requires no measurable
computational effort from the attacker and includes traditional DoS
attacks such as pipe stoppage.  An effortful
attack requires the attacker to invest in the system and therefore
requires computational effort.

\subsection{Security Goals}
\label{sec:attritionAnalysis:securityGoals}

The overall goal of the LOCKSS system is to maintain a high probability
that the consensus of peers reflects the correct AU, and a high
probability that a reader accesses good data.  In contrast, an attrition
adversary's goal is to decrease these probabilities significantly by
preventing peers from auditing their replicas for a long time, long
enough for undetected storage problems such as ``bit rot'' to occur.

Severe pipe stoppage attacks in the wild last for days or
weeks~\cite{Moore2001short}.  Our goal is to ensure that, in the very
least, the LOCKSS system withstands such attacks sustained over months.
Beyond pipe stoppage, attackers must use protocol messages to some
extent.  We seek to ensure the following three conditions.  First, a peer manages
its resources so as to prevent exhaustion no matter how much effort is
exerted by however many identities request service.  Second, when
deciding which requests to service, a peer gives preference to requests
from those likely to behave properly (i.e., ``ostensibly legitimate'').
And third, at every stage of a protocol exchange, an ostensibly
legitimate attacker expends commensurate effort to that which he imposes
upon the defenders.

\subsection{Defensive Framework}
\label{sec:attritionAnalysis:defensiveFramework}

We seek to curb the adversary's success by modeling a peer's processing
of inbound messages as a series of filters, each costing a certain
amount to apply.  A message rejected by a filter has no further effect
on the peer, allowing us to estimate the cost of eliminating whole
classes of messages from further consideration.  Each filter increases
the effort a victim needs to defend itself, but
limits the effectiveness of some adversary capability.

The \emph{bandwidth filter} models a peer's network connection. It
represents the physical limits on the rates of inbound messages that an
adversary can force upon his victims.  The \emph{admission control
filter} takes inbound messages at the maximum rate supported by the
bandwidth filter and further limits them to match the maximum rate at
which a peer expects protocol traffic from legitimate senders,
favoring known peer identities.  This curbs
the adversary's use of unlimited identities and prevents him from
applying potentially unconstrained computational resources upon a
victim.  The \emph{effort balancing filters} ensure that effort
imposed upon a victim by ostensibly legitimate traffic is balanced by
correspondingly high effort borne by the attacker, making it costly for
a resource-constrained adversary to masquerade as a legitimate peer.

We show in Section~\ref{sec:effortful:poll} that the most effective strategy
for effortful attacks is to emulate legitimacy,
and that even this
has minimal effect on the utility of the system.  Effortless attacks,
such as traditional distributed DoS (DDoS) attacks, are more effective but
must be maintained for a long time against most of the peer
population to
degrade the system significantly (Section~\ref{sec:effortless:pipe}).

\section{The LOCKSS Replica Auditing and Repair Protocol}
\label{sec:protocol:lockss}

The LOCKSS audit process operates as a sequence of ``opinion polls''
conducted by every peer on each of its AU replicas.  At intervals,
typically every 3 months, a peer (the \emph{poller}) constructs a
random subset (i.e., sample) of the peer population that it knows are
preserving an AU, and invites those peers as \emph{voters} into a poll.
Each voter individually hashes a poller-supplied nonce and its replica
of the AU to produce a fresh vote, which the poller tallies.  If the
poller is outvoted in a landslide (e.g., it disagrees with 80\% of the
votes), it assumes its replica is corrupt and repairs it from a
disagreeing voter.  The roles of poller and voter are distinct, but
every peer plays both.

\begin{figure}
\centerline{\includegraphics{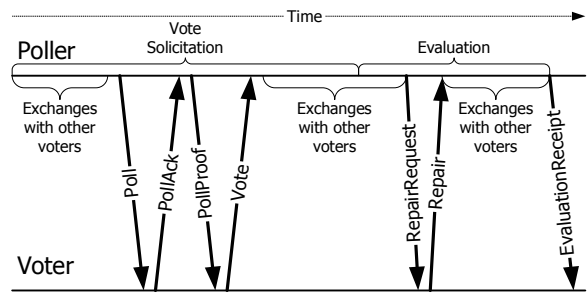}}
\caption{\label{fig:protocolOverview} A time-line of a poll,
showing the message exchange between the poller and a voter.
}
\end{figure}

The general structure of a poll follows the time-line of
Figure~\ref{fig:protocolOverview}. A poll consists of two phases: the
\emph{vote solicitation} phase and the \emph{evaluation} phase.
In the vote solicitation phase the poller requests and obtains
votes from as many voters in its sample of the population as possible.
Then the poller begins the evaluation phase,
during which it compares these votes to its own replica,
one hashed content block at a time,
and tallies them.
If the hashes disagree the poller may request repair blocks from its
voters and re-evaluate the block.
If in the eventual tally, after any repairs, the poller agrees with the landslide majority,
it sends a receipt to each of its voters and immediately starts a new
poll.
Peers interleave making progress on their own polls  and voting in other
peers' polls,  spreading each poll over a long period chosen so
that polls on a given AU occur at a rate much higher than
the rate of undetected storage problems, e.g. ``bit rot.''

\subsection{Vote Solicitation}
\label{sec:protocol:solicitation}

The outcome of a poll is determined by the votes of the \emph{inner circle}
peers,  sampled at the start of the poll by the poller from its
\emph{reference list} for the AU.
The reference list contains mostly peers that have agreed with the poller in
recent polls on the AU, and a few peers from its static \emph{friends list},
maintained by the poller's operator.

A poll is considered successful if its result is based on a minimum
number of inner circle votes, the \emph{quorum}, which is
typically 10, but may change according to the application's needs for
fault tolerance. To ensure that a poll is likely to succeed, a poller
invites into its poll a larger inner circle than the quorum (typically,
twice as large).  If at first try, an inner circle peer fails to
respond to an invitation, or refuses it, the poller
contacts a different inner circle voter,
re-trying the reluctant peer later in the same vote solicitation phase.

An individual vote solicitation consists of four messages (see
Figure~\ref{fig:protocolOverview}): \msg{Poll}, \msg{PollAck},
\msg{PollProof}, and \msg{Vote}.  For the duration of a poll, a
poller establishes an encrypted TLS session with each voter
individually, via an anonymous Diffie-Hellman key exchange. Every
protocol message is conveyed over this TLS session, either keeping
the same TCP connection from message to message, or resuming the
TLS session over a new one.

The \msg{Poll} message invites a voter to participate in a poll on an 
AU. The invited peer
responds with a \msg{PollAck} message, indicating either a refusal to
participate in the poll at the time, or an acceptance
of the invitation, if it can compute a vote within a predetermined time
allowance.  The voter commits and reserves local resources to that
effect.  The \msg{PollProof} message supplies the voter with a random
nonce to be used during vote construction.  To
compute its vote, the voter uses a cryptographic hash function (e.g.,
SHA-1) to hash the nonce supplied by the poller,
followed by its replica of the AU, block by block.  The vote consists of the running hashes
produced at each block boundary.  Finally, the voter sends its vote back
to the poller in a \msg{Vote} message.

These messages also contain proofs of computational
effort, such as those introduced by Dwork et al.~\cite{Dwork1992short},
sufficient to ensure that, at every protocol stage, the requester of a service
has more invested in the exchange than the supplier of the service (see
Section~\ref{sec:defenses:admissionControl}).

\subsection{Peer Discovery}
\label{sec:protocol:discovery}

The poller uses the vote solicitation phase of a poll not only to obtain
votes for the current poll, but also to discover new peers for its
reference list from which it can solicit inner circle votes in
future polls.

Discovery is effected via \emph{nominations} included in 
\msg{Vote} messages.  A voter
picks a random subset of its current reference list, which it
includes in the \msg{Vote} message.  The poller
accumulates these nominations.  When it concludes 
its inner circle solicitations, it chooses a random sample of these nominations
as its \emph{outer circle}.  It proceeds to
solicit regular votes from these outer circle peers in a manner identical
to that used for inner circle peers.

The purpose of the votes obtained from outer circle voters is to show
the ``good behavior'' of newly discovered peers.  Those who perform
correctly, by supplying votes that agree with the prevailing outcome
of the poll, are added into the poller's reference list at the
conclusion of the poll; the outcome of the poll is
computed only from inner-circle votes.

\subsection{Vote Evaluation}
\label{sec:protocol:evaluation}
Once the poller has accumulated all votes it could obtain from
inner and outer circle voters, it begins the poll's evaluation phase.
During this phase, the poller computes, in parallel, all block
hashes that each voter \emph{should have} computed, if that voter's replica
agreed with the poller's.  A vote \emph{agrees} with the
poller on a block if the hash in the vote and that computed by the
poller are the same.

For each hash computed by the poller for an AU block, there are three
possibilities: first, the landslide majority of inner-circle votes
(e.g., 80\%) agree with the poller; in this case, the poller considers
the audit successful up to this block and proceeds with the next block.
Second, the landslide majority of inner-circle votes disagree with the
poller; in this case, the poller regards its own replica of the AU as
damaged, obtains a repair from one of the
disagreeing voters (via the \msg{RepairRequest} and \msg{Repair}
messages), and reevaluates the block hoping to find itself in the
landslide majority, as above.  Third, if there is no landslide majority of
agreeing or disagreeing votes, the
poller deems the poll inconclusive, raising an alarm that requires
attention from a human operator.

Throughout the evaluation phase, the poller may also decide to obtain a
repair from a random voter, even if one is not required (i.e., even if
the corresponding block met with a landslide agreement). The purpose of such
\emph{frivolous} repairs is to prevent targeted free-riding via the
refusal of repairs; voters are expected to supply a small number of
repairs once they commit to participate in a poll, and are penalized
otherwise (Section~\ref{sec:defenses:admissionControl}). 

If the poller hashes all AU blocks without raising an alarm, it
concludes the poll by sending an evaluation receipt to each
voter (with an \msg{EvaluationReceipt} message), indicating that it
will not be requesting any more repairs.  The poller then updates its
reference list by removing all voters whose votes determined the poll
outcome and by inserting all agreeing outer-circle voters and some
peers from the friends list (for details
see~\cite{Maniatis2003lockssSOSPshort}).
The poller then restarts a
poll on the same AU, scheduling it to conclude an inter-poll interval
into the future.

\section{LOCKSS Defenses}
\label{sec:defenses}

Here we outline the attrition defenses of the LOCKSS protocol: admission
control, desynchronization, and redundancy.  These defenses raise system
costs for both
loyal peers and attackers, but favor ostensible legitimacy.
Given a constant amount of over-provisioning, loyal peers continue
to operate at the necessary rate regardless of the attacker's power.
Many systems over-provision resources to protect performance from
known worst-case behavior (e.g., 
the Unix file system~\cite{McKusick84short}).
 

\subsection{Admission Control}
\label{sec:defenses:admissionControl}

The purpose of the admission control defense is to ensure that a peer
can control the rate at which it considers poll invitations from
others, favoring invitations from those who operate at roughly the same
rate as itself and penalizing others.
We implement admission control using three mechanisms: rate limitation,
first-hand reputation, and effort balancing.


{\bf Rate Limitation}: Without limits on the rate at which they attempt
to service requests,
peers can be overwhelmed by floods of ostensibly valid requests.
\emph{Rate Limitation} suggests that peers should initiate and
satisfy requests \emph{no faster than necessary} rather than \emph{as
fast as possible}.  Because readers access only their local LOCKSS peer,
the audit and repair protocol is not subject to end-users'
unpredictable request patterns.  The protocol can proceed at its own pace,
providing an interesting test case for rate limitation.

We identify three possible attacks based on deviation from the
\emph{necessary} rate of polling.
A \emph{poll rate} adversary would seek to trick victims into either
decreasing (e.g., by causing back-off behavior) or increasing (e.g.,
in an attempt to recover from a failed poll) their rate of calling
polls.
A \emph{poll flood} adversary would seek,
under a multitude of identities, to invite victims into as many frivolous
polls as possible hoping to crowd out the legitimate poll requests
and thereby reduce the ability of loyal peers to audit and repair their
content. 
A \emph{vote flood} adversary would seek to supply as many bogus votes
as possible hoping to exhaust loyal pollers' resources in useless
but expensive proofs of invalidity.

Peers defend against all these adversaries by
setting their rate limits autonomously, not varying them in response
to other peers' actions.
Responding to adversity (inquorate polls or perceived contention)
by calling polls more frequently could aggravate the problem;
backing off to a lower rate of polls would achieve the adversary's
aim of slowing the detection and repair of damage; Kuzmanovic et
al.~\cite{Kuzmanovic2003short} describe a similar attack in the context
of TCP retransmission timers.
Because peers do not react, the \emph{poll rate} adversary has no
opportunity to attack.  The price of this fixed rate of operation is that, 
absent manual intervention, a peer may
take several inter-poll intervals to recover from a catastrophic storage
failure.

The \emph{poll flood} adversary tries to get victims to over-commit their
resources or at least to commit excessively to the adversary.
To prevent over-commitment, peers maintain a task schedule of their
promises to perform effort, both to generate votes for others and
to call their own polls.  If the effort of computing the
vote solicited by an incoming \msg{Poll} message
cannot be accommodated in the schedule, the invitation is refused.
Furthermore, peers limit the rate at which they even \emph{consider} poll
invitations (i.e., establishing a secure session, checking their
schedule, etc.).  A peer sets this rate limit for considering
poll invitations according to the rate of poll
invitations it sends out to others; this is essentially a
\emph{self-clocking} mechanism.  We explain how this rate limit is enforced
in the first-hand reputation description below.
We evaluate our defenses against poll flood strategies in
Section~\ref{sec:effortless:admissionControl}.


The \emph{vote flood} adversary is hamstrung by the fact that votes
can be supplied only in response to an invitation by the putative
victim poller, and pollers solicit votes at a fixed rate.  Unsolicited votes are ignored.

{\bf First-hand reputation}: A peer locally maintains and uses 
first-hand reputation (i.e., history) for other peers.  Each peer $P$
maintains a \emph{known-peers} list, separately for each AU it
preserves. The list contains an entry for every peer that $P$ has encountered in the
past, tracking its exchange of votes with that peer.
The entry holds a reputation grade for the peer, which is one of
three values: \emph{debt}, \emph{even}, or \emph{credit}.  A debt
grade means that the peer has supplied $P$ with fewer votes than $P$
has supplied it.  A credit grade means $P$ has supplied the peer with
fewer votes than the peer has supplied $P$.  An even grade means that
$P$ and the peer are even in their recent exchanges of votes.
Entries in the known-peers list ``decay'' with time toward the
\emph{debt} grade.

In a protocol interaction, both the poller and a voter modify
the grade they have assigned to each other depending on their
respective behaviors.  If the voter supplies a valid vote and valid repairs
for any blocks the poller requests, then the poller increases the grade it has
assigned to the voter (from debt to even, from even to credit, or from
credit to credit) and the voter correspondingly decreases the grade it
has assigned to the
poller.  If either
the poller or the voter misbehave (e.g., the voter commits to supplying a
vote but does
not, or the poller does not send a valid evaluation receipt), then the other
peer decreases its grade to debt.
This is similar to the reciprocative strategy of Feldman et al.~\cite{Feldman2004short},
in that it penalizes peers who do not reciprocate, i.e., do not supply
votes in return for the votes they receive.  

Peers randomly drop some poll invitations arriving from previously
unknown peers and from pollers with a debt grade.
Invitations from pollers with an even or credit grade are not dropped.
This reputation system
reduces free-riding, as it is not possible for a peer to maintain an even
or credit grade without providing valid votes.
To discourage identity whitewashing the drop probability imposed on unknown
pollers is higher than that imposed on known in-debt pollers.
Furthermore, invitations from unknown or in-debt pollers are subject to a
rigid rate limit; after it admits one such invitation for
consideration, a voter enters a \emph{refractory} period during which
it automatically rejects all invitations from unknown or in-debt pollers.
Like the known-peers list, refractory periods are maintained on a per AU
basis.  
Consequently, during every refractory period, a voter admits at most one
invitation from unknown or in-debt peers, plus at most one invitation from
each of its fellow peers with a credit or even grade.  Since credit and even
grades decay with time, the total ``liability'' of a peer in the number
of invitations it must admit per refractory period is limited to a small
constant number.  As a result, the duration of the
refractory period is inversely proportional to the rate limit imposed by
the peer on the poll invitations that it considers for each AU it preserves.

Continuous triggering of the refractory period can stop a victim voter
from accepting invitations from unknown peers who are loyal; this can
limit the choices a poller has in potential voters to peers that know
the poller already.  To reduce this impediment to diversity, we
institute the concept of peer \emph{introductions}.  A peer may
introduce peers that it considers loyal to others; 
peers introduced in this way can bypass random drops and refractory
periods.  Introductions are bundled along with nominations during
the regular discovery process (Section~\ref{sec:protocol:discovery}).
Specifically, a poller randomly partitions the peer identities in a 
\msg{Vote} message into outer circle nominations and
introductions.
A poll invitation from an introduced peer is
treated as if coming from a known peer with
an even grade.  This unobstructed
admission consumes the introduction in such a way that at
most one introduction is honored per (validly voting) introducer, and
unused introductions do not accumulate.  Specifically, when consuming the
introduction of peer $B$ by peer $A$ for AU $X$, 
all other introductions of other
introducees by
peer $A$ for AU $X$ 
are ``forgotten,'' as are all introductions of peer $B$ for $X$ by
other introducers.  Furthermore, introductions by peers who have
entered and left the reference list are also removed, and the maximum
number of outstanding introductions is capped.


{\bf Effort Balancing:} If a peer expends more effort to react to
a protocol message than the sender of that message did to generate and
transmit it, then an attrition attack need consist only of a flow of
ostensibly valid protocol messages, enough to exhaust the victim peer's
resources.

We adapt the ideas of pricing via processing~\cite{Dwork1992short}
to discourage such attacks from resource-constrained adversaries
by \emph{effort balancing} our protocol.
We inflate the cost of a request by requiring it to include
a proof of computational effort sufficient to ensure that the
total cost of generating the request exceeds the cost to the
supplier of both verifying the effort proof and satisfying the request.
We favor Memory-Bound Functions
(MBF)~\cite{Dwork2003short} rather than CPU-bound schemes
such as ``client puzzles''~\cite{Dean2001short} for this purpose, because the
spread in memory system performance is smaller than that
of CPU performance~\cite{Douceur2002short}.  Note that
an adversary with ample computational resources is not
hindered by effort balancing.

Applying an effort balancing filter at each step of a multi-step protocol
defends
against three attack patterns: first, \emph{desertion} strategies in
which the attacker stops taking part some way through the protocol, having
spent less effort in the process than the effort inflicted upon his
victim; second, \emph{reservation} strategies that cause the victim to
schedule or commit resources that the attacker does not use,
but successfully take away from other, useful tasks; and, third, \emph{wasteful} strategies
in which service is obtained but the result is not ``consumed'' by the
requester as expected by the
protocol, in an attempt to minimize the attacker's total expended effort.

Pollers could mount a desertion attack by cheaply soliciting
an expensive vote.
To discourage this, the poller must include
provable effort in its vote solicitation
messages (\msg{Poll} and \msg{PollProof}) that in total
exceeds,
by at least an amount described in the next paragraph,
the effort required by the voter to verify that effort
and to produce the requested vote.
Producing a vote amounts to
fetching an AU replica from disk, hashing it, and shipping back to the
poller one hash per block in the \msg{Vote} message.

Voters could mount a desertion attack by cheaply generating a bogus
vote in response to an expensive solicitation,
returning garbage instead of block hashes in the hope
of wasting not merely the poller's solicitation effort but also its effort
to verify the hashes.
Because the poller evaluates the vote one block at a time, it
costs the effort of hashing one block to detect that the
vote disagrees with its own AU replica, which may mean
either that the vote is bogus, or that the poller's and voter's
replicas of the AU differ in that block.
The voter must therefore include in the \msg{Vote} message
provable effort sufficient to cover the cost of hashing a
single block and of verifying this effort.
The extra effort in the solicitation messages referred to above
is required to cover the generation of this provable effort.

Pollers could mount a reservation attack by sending a valid \msg{Poll}
message that causes the voter to reserve time for computing a vote
in anticipation of a \msg{PollProof} message which the poller
never sends.
To discourage this,  pollers must include sufficient
\emph{introductory effort} in \msg{Poll} messages
to match the effort the voter
\emph{could} expend while waiting for the poller's \msg{PollProof} before
timing out.
Upon receiving the affirmative \msg{PollAck}, the poller performs the
balance of the provable effort the voter needs to defend against desertion
attacks. The poller includes the resulting proof in the \msg{PollProof} message.

Pollers could mount a wasteful attack by soliciting expensive
votes and then discarding them unevaluated.
To discourage this we require the poller,
after evaluating a vote,
to supply the voter with an unforgeable \emph{evaluation receipt}
proving that it evaluated the vote.
Voters generate votes and pollers evaluate them using very
similar processes: generating or validating effort proofs and hashing
blocks of the local AU replica.  Conveniently, generating a proof
of effort using our chosen MBF mechanism also generates
160 bits of unforgeable byproduct.  The voter remembers the byproduct;
the poller uses it as the evaluation receipt to send to the voter.
If the receipt matches the voter's remembered byproduct the voter
knows the poller performed the necessary effort, regardless of whether the
poller was loyal or malicious.

In Section~\ref{sec:effortful:poll} we show how our series of
effort balancing filters fare against such attacks mounted 
by pollers, evaluating the success of an adversary who defects at
different key points
in the protocol, seeking to maximize the defenders' wasted effort.
We omit the evaluation of effort balancing
attacks by voters, since they are rendered ineffective by the rate
limits described above.


\subsection{Desynchronization}
\label{sec:defenses:desynchronization}

The \emph{desynchronization} defense requires that measures
such as randomization
be applied to avoid the kind of
inadvertent synchronization that has been observed in many distributed systems.
Examples include TCP sender windows at bottleneck routers,
clients waiting for a busy server,
and periodic routing messages~\cite{Floyd1994short}.
Peer-to-peer systems in which a peer requesting service must find
many others simultaneously available to supply that service
(e.g., in a read-one-write-many fault-tolerant
system~\cite{Malkhi1998short})
may encounter this problem.
If they do, even
absent an attack, moderate levels of peer busyness can prevent the
system from delivering services.  A poll flood attacker in this situation may
only need to increase peer busyness slightly to have a large effect.

Simulations of poll flood attacks on an earlier version of the
protocol~\cite{Maniatis2004short} showed this effect.
Loyal pollers were at a great disadvantage against
the attrition adversary because they needed to find a quorum of
voters who could simultaneously vote on an AU.
The voters must be chosen at random to make directed subversion hard
for the adversary. They must also have free resources at the specified
time, in the face of resource contention from other peers who are also
competing for voters on the same or other AUs at the same time.
The adversary has no such requirements;
he can find and invite an individual victim into a futile poll.

Peers avoid this problem
by soliciting votes individually rather
than synchronously, extending the period during which a quorum
of votes can be collected before they are all evaluated.
A poll is thus a sequence of two-party interactions rather than
a single multi-party interaction.


\subsection{Redundancy}
\label{sec:defenses:redundancy}

If the survival of, or access to, an AU relied only on a few replicas,
an attrition attack could focus on those replicas, cutting off the
communication between them needed for audit and repair.
Each LOCKSS peer preserving an AU maintains its own replica and serves it
only to its local clients.  This massive redundancy helps resist
attacks in two ways.
First, it ensures that a successful attrition attack must target most
of the replicas,  typically a large number of peers.
Second, it forces the attrition attack to suppress the communication
or activity of the targeted peers continuously for a long period.
Unless the attack does both, the targeted peers recover by
auditing and repairing themselves from the untargeted peers, as shown in
Section~\ref{sec:effortless:pipe}.
This is because massive redundancy allows peers at each poll
to choose a sample of their reference list that is bigger
than the quorum and continue to solicit votes from them
at random times for the entire duration of a poll (typically 3 months) until
the voters accept.
Further, the margin between the rate at which
peers call polls and the rate at which they suffer undetected damage provides
redundancy in time.  A single failed poll has little effect on the safety
of its caller's replica.

\section{Simulation}
\label{sec:simulation}

In this section we give details about the simulation environment and the
metrics we use to evaluate the system's effectiveness in meeting its goals.

The first requirement for the system is to preserve the long-term integrity
of the replicated AU, by ensuring that a
majority of all replicas reflect the correct AU contents.
If a majority of replicas are corrupt, we consider the system to
have failed with irrecoverable damage.

An attrition adversary could in theory mount a pipe stoppage
attack of sufficient coverage, intensity, and duration to prevent
all communication between all replicas for long enough for undetected
storage errors
to corrupt a majority of replicas. In practice this attack would have to last
for years;
other non-attrition attacks aimed more directly at corrupting data are
more likely to make progress at these timescales~\cite{Maniatis2003lockssSOSPshort}.

The second requirement for the system is to
preserve access to a correct replica at each peer for as much of
the time as possible, in the face of local storage failures and attacks.
Reducing the probability that a particular correct replica is accessible is
a more attainable goal for an attrition attack than causing irrecoverable
damage through an attrition attack, thus our simulations measure the success of the adversary
against this goal.

\subsection{Evaluation Metrics}
\label{sec:simulation:evaluation}

We use four metrics to measure the effectiveness of our defenses
against the attrition adversary: 
access failure probability, delay ratio,
coefficient of friction, and cost ratio.

\emph{Access failure probability}:  To measure the success of an
attrition adversary at increasing the probability that a reader obtains
a damaged AU replica, we compute the access failure probability as
the fraction of all replicas in the system that are damaged, averaged over
all time points in the experiment.

\emph{Delay ratio}: To measure the degradation an attrition adversary
achieves, we compute the delay ratio as the mean time between successful polls at loyal
peers with the system under attack divided by the same measurement without
the attack.

\emph{Coefficient of friction}: To measure the cost of an attack to loyal
peers, we measure the coefficient of friction, defined as the
average effort expended by loyal peers per successful poll during an attack divided by their
average per-poll effort absent an attack.

\emph{Cost ratio}: To compare the cost of an effortful attack to the
adversary and to the defenders, we compute the cost ratio, which is the ratio of the
total effort expended by the attackers during an attack to that of the defenders.

\subsection{Environment and Adversaries}

We run our experiments using Narses~\cite{Giuli2002},
a discrete-event simulator that exports a sockets-like network interface
and provides facilities for modeling computationally expensive operations,
such as computing MBF efforts and
hashing documents.  Narses allows experimenters to pick from a range of
network models that trade off speed for accuracy.  In this study we are
mostly interested in the application-level effects of an attrition
attack, so we choose a simplistic network model that takes into
account network delays but not congestion, except for the side-effects
of artificial congestion used by a pipe stoppage adversary.
The link bandwidths with which peers connect to the network are
uniformly distributed
among three choices: 1.5, 10, and 100 Mbps.  Link latencies are
uniformly distributed between 1 and 30 milliseconds.

Nodes in the system are divided into two categories: loyal
peers and the adversary's minions.  Loyal peers are uncompromised peers that execute
the protocol correctly.  Adversary minions are nodes that collaborate to
execute the adversary's attack strategy.

We conservatively simulate the adversary as a cluster of
nodes with as many IP addresses and as much compute power as he needs.
Each adversary minion has complete and instantaneous knowledge of all
adversary state and has a magically incorruptible copy of all AUs.
Other assumptions about our adversary that are less relevant to attrition can be
found in~\cite{Maniatis2003lockssSOSPshort}.


To distill the adversary's actual cost of attack from any other effort he might
have to shoulder (e.g., to masquerade as a loyal peer), the adversary in these
experiments is completely outside of the network of loyal peers.
Loyal peers never ask the adversary's minions to vote in polls and the
adversary only asks loyal peers to vote in his polls.  This differs
from LOCKSS adversaries we have studied before~\cite{Maniatis2003lockssSOSPshort}.


\subsection{Simulation Parameters}
\label{sec:simulation:parameters}

We evaluate the preservation of a collection of AUs distributed among a
population of loyal peers.  For simplicity in this stage of our
exploration, we assume that each AU contains 0.5 GBytes (a large AU in
practice).  Each peer maintains a number of AUs ranging from 50 to
600. All peers have replicas of all AUs; we do not yet simulate the
diversity of local collections that we expect will evolve over time.
These simplifications allow us to focus our attention on the common
performance of our attrition resistance machinery, ignoring for the time
being how that performance varies when AUs vary in size and
popularity.
Note that our 600 simulated AUs total about 10\% of the
size of the annual AU intake of a large journal collection such
as that of Stanford University
Libraries. Adding the equivalent of 10 of today's low-cost PCs per year
and consolidating them as old
PCs are rendered obsolete is an affordable
deployment scenario for a large library.
We set
all costs of primitive operations (hashing, encryption, L1 cache and RAM
accesses, etc.) to match the capabilities of such a low-cost PC.

All simulations have a constant loyal peer population of 100 nodes and
run for two simulated years, 
with 3 runs per data point.  Each peer runs
a poll on each of its AUs on average every 3 months. Each poll uses a
quorum of 10 peers and considers landslide agreement as having a maximum
of three disagreeing votes.  These parameters were empirically
determined from previous iterations of the deployed beta protocol.  We
set the fixed drop probability to be 0.90 for unknown peers and 0.80 for
in-debt peers.

Memory limits in the Java Virtual Machine prevent Narses from simulating
more than about 50 AUs/peer in a single run. We simulate 600 AU collections
by \emph{layering} 50 AUs/peer runs,
adding the tasks caused by this layer's 50 AUs to the
task schedule for each peer accumulated during the preceding
layers.
In effect,  layer $n$ is a simulation of 50 AUs on peers already
running a realistic workload of $50(n - 1)$ AUs.
The effect is to over-estimate the peer's busyness for AUs in higher
layers and under-estimate it for AUs in lower layers;
AUs in a layer compete for the resources left over by lower layers
but AUs in lower layers are unaffected by the resources used in
higher layers.
We have validated this technique against unlayered simulations in
smaller collections, as well as against simulations in which inflated per-AU preservation
costs cause similar levels of peer load; we found negligible differences.


We are currently exploring the parameter space but use the following
heuristics to help determine parameter values.  The
refractory period of one day
allows for 90 invitations from unknown or in-debt peers to be accepted 
per 3-month inter-poll interval; in contrast, a peer requires an
average of 30 votes per poll and, because of self-clocking,
should be able to accept at least an average of 30 poll
invitations per inter-poll interval. Consequently, we allow up to a
total of four times the rate of poll invitations that should be expected
in the absence of attacks.  At this rate, even if all poll invitations
are bogus, the total cost of detecting them as bogus is negligible.

We set the fixed drop probability for in-debt peers and the cost of verifying
an introductory effort so that the cumulative introductory effort
expended by an effortful
attack on dropped invitations is more than the
voter's effort to consider the adversary's eventually \emph{admitted}
invitation.  Since an adversary has to try with in-debt identities on
average 5 times to be admitted (thanks to the $1-0.8=0.2$ admission
probability), we set the introductory effort to be 20\% of the total
effort required of a poller; by the time the adversary has gotten his
poll invitation admitted, even if he defects for the rest of the poll,
he has already expended on average 100\% of the effort he would have, had he
behaved well in the first place.

\section{Results}
\label{sec:Results}

\begin{figure}
\centerline{\includegraphics{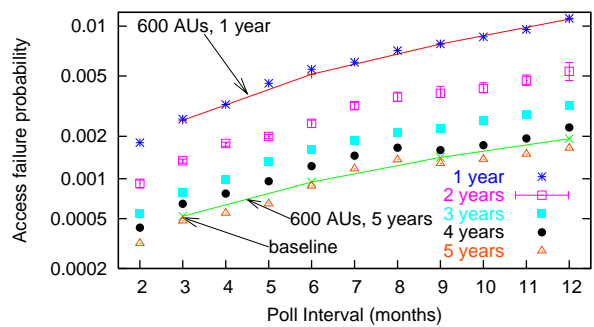}}
\caption{Mean access failure probability ($y$ axis in
log scale) for increasing inter-poll intervals ($x$ axis)
at variable mean times between storage failure (from 1 to 5 years
per disk), absent an attack.  We show results for
collection sizes of 50 (points only) and of 600 AUs (linespoints).
For illustration, we show minimum and maximum values for the
2-year data set; this variance is representative of all measurements,
which we omit for clarity.}
\label{fig:calibration:damageRateCombined}
\end{figure}

The probability of access failure summarizes the success of an
attrition attack.
We start by establishing a baseline rate of
access failures absent an attack.
We then assess the effectiveness against this baseline of the
effortless attacks we consider:
network-level flooding attacks on the bandwidth filter in Section~\ref{sec:effortless:pipe},
and Sybil attacks on the admission control filter in Section~\ref{sec:effortless:admissionControl}.
Finally,
we assess against this baseline each of the effortful attacks
corresponding to each effort verification filter in Section~\ref{sec:effortful:poll}.

In each case we show the effect of increasing scales of attack on
the access failure probability, and relevant supporting graphs including
the delay ratio, the coefficient of friction, and for effortful attacks the
cost ratio.

Our mechanisms for defending against an attrition adversary raise
the effort required per loyal peer.
To achieve a bound on access failure probabilities, one must
be willing to over-provision the system to accommodate the extra effort.
Over-provisioning the system by a
constant factor, we can defend it against application-level attrition attacks
of unlimited power (Sections~\ref{sec:effortless:admissionControl} and
\ref{sec:effortful:poll}).

\subsection{Baseline}
\label{sec:baseline}

Our simulated peers suffer random
storage damage at rates of one block in 1 to 5 disk years (50 AUs per disk). This is
a very high rate of damage, as the
LOCKSS beta test suggests that one such \emph{undetected} occurrence
every 5 machine years would be a gross overestimate; we inflate this
failure rate to encompass other types of storage failure, including temporary
tampering and human error.
Figure~\ref{fig:calibration:damageRateCombined} plots the access failure
probability versus the inter-poll interval. It shows that as the
inter-poll
interval increases relative to the mean interval between storage
failures, the access failure probability increases because damage
takes longer to detect and repair.  The access failure probability is
similar for a 50-AU collection all the way up to a 600-AU collection
(we omit intermediate collection sizes for clarity).

For comparison purposes in the rest of the experiments, the baseline
access failure probability of $4.8\times 10^{-4}$ for a 50-AU collection
and of $5.2\times 10^{-4}$ for a 600-AU collection correspond to our
inter-poll interval of 3 months and a storage damage rate of one block
per 5 disk years.

\begin{figure}
\centerline{\includegraphics{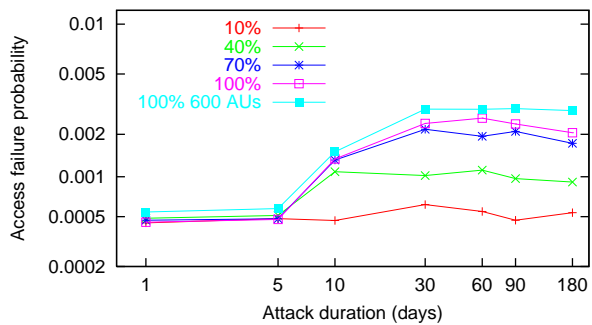}}
\caption{The access failure probability ($y$ axis in log scale) observed
during repeated pipe stoppage attacks of varying duration ($x$ axis in log
scale), covering between 10 and 100\% of the peers.}
\label{fig:effortless:pipeStoppageAccessFailure}
\end{figure}

\subsection{Targeting the Bandwidth Filter}
\label{sec:effortless:pipe}

The ``pipe stoppage'' adversary models packet flooding or more
sophisticated attacks~\cite{Kuzmanovic2003short}.  This adversary
suppresses all communication between some proportion of the total peer
population (its \emph{coverage}) and other
LOCKSS peers.  During a
pipe-stoppage attack, local readers may still access content.  
Each attack consists
of a period of pipe stoppage, which lasts between 1 and 180 days,
followed by a 30-day recuperation period
during which all communication is restored; this pattern is repeated for the
entire experiment, affecting a different random subset of
the population in each iteration.

Figure~\ref{fig:effortless:pipeStoppageAccessFailure} plots the access
failure probability versus the attack duration for varying coverage
values (10 to 100\%).  As expected, the access failure probability
increases as the coverage of the attack increases.
At 100\% coverage, we see that the larger 600-AU collection tracks
the small 50-AU collection closely, albeit at a slight disadvantage, due
to increased scheduling contention as peers get more loaded.
Even in
the extreme case where 100\% of the population has no communication for
6 months,
access failures occur with a mean probability of about $2.9 \times
10^{-3}$ for a 600-AU collection; this is well within tolerable limits
for any service that is
widely open to the Internet.

\begin{figure}
\centerline{\includegraphics{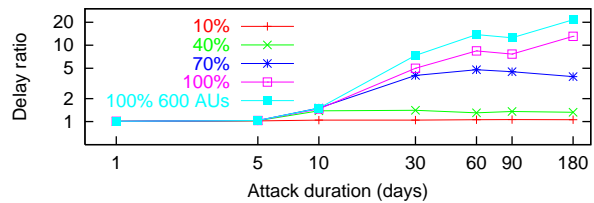}}
\caption{The delay ratio ($y$ axis in log scale) imposed by repeated pipe
stoppage attacks of varying duration ($x$ axis in log scale)
and coverage of the population.  Absent an
attack, this metric has value 1.}
\label{fig:effortless:pipeStoppageDelay}
\end{figure}

\begin{figure}
\centerline{\includegraphics{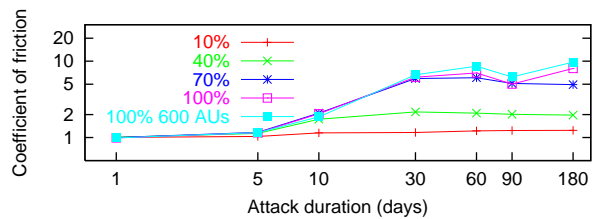}}
\caption{The coefficient of friction ($y$ axis in log scale) imposed by
pipe stoppage attacks of varying duration ($x$ axis in log scale)
and coverage of the population.}
\label{fig:effortless:pipeStoppageCost}
\end{figure}

Figures~\ref{fig:effortless:pipeStoppageDelay}
and~\ref{fig:effortless:pipeStoppageCost} plot the delay ratio and
coefficient of friction, respectively, versus attack duration.  
We find that attacks must last at least 60 days to
raise the delay ratio by an order of magnitude.  Similarly, the
coefficient of friction during repeated attacks that last less than a
few days each is negligibly greater than 1; for longer attacks, the
coefficient can reach 10.

\begin{figure}
\centerline{\includegraphics{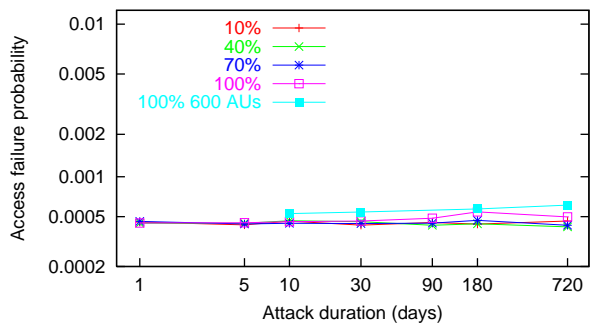}}
\caption{The access failure probability ($y$ axis in log scale) for
attacks of increasing duration ($x$ axis in log scale) by the admission
control adversary over 10 to 100\% of the peer population.}
\label{fig:effortless:sybil:accessFailure}
\end{figure}

\subsection{Targeting the Admission Control Filter}
\label{sec:effortless:admissionControl}

The admission control adversary aims to reduce the likelihood of a
victim admitting a loyal
poll request by triggering that victim's refractory
period as often as possible.  This adversary sends cheap garbage
invitations to varying fractions of the  peer population for varying periods of
time separated by a fixed recuperation period of 30 days. The adversary
sends his invitations using poller addresses that are unknown to the
victims.  These, when eventually admitted, cause those victims to enter
their refractory periods and drop all subsequent invitations from 
unknown and in-debt peers.

\begin{figure}
\centerline{\includegraphics{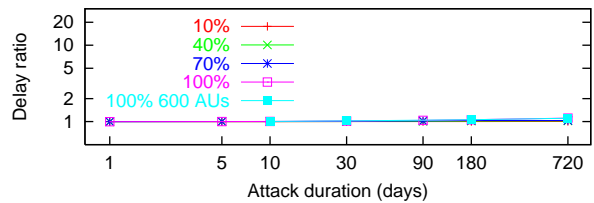}}
\caption{Delay ratio ($y$ axis in log scale)
for attacks of increasing duration ($x$ axis in log scale) by the
admission control adversary over 10 to 100\% of the peer population.}
\label{fig:effortless:sybil:delayratio}
\end{figure}

\begin{figure}
\centerline{\includegraphics{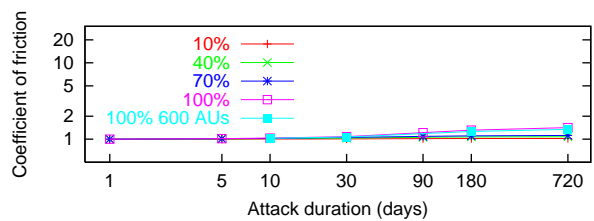}}
\caption{Coefficient of friction ($y$ axis in log scale)
for attacks of increasing duration ($x$ axis in log scale) by the
admission control adversary over 10 to 100\% of the peer population.}
\label{fig:effortless:sybil:cf}
\end{figure}

Figures~\ref{fig:effortless:sybil:accessFailure}, 
\ref{fig:effortless:sybil:delayratio}, and~\ref{fig:effortless:sybil:cf}
show that these attacks have
little effect on the access failure probability or the delay ratio.  The
access failure probability is raised to $5.9 \times 10^{-4}$ when
the duration of the attack reaches the entire duration of our
simulations (2 years) for full population coverage and a 600-AU collection.  At that attack
intensity, loyal peers no longer admit poll invitations from unknown or
in-debt loyal peers, unless supported by an introduction.  This causes
discovery to operate more slowly; loyal peers waste a lot of their
resources on introductory effort proofs that are summarily rejected by
peers in their refractory period.  This wasted effort is apparent in
Figure~\ref{fig:effortless:sybil:cf}, which shows that when the attack
is sustained for long periods of time, it
can raise the cost to loyal peers of every successful poll by 33\%.

Note that techniques such as blacklisting,
commonly used to defeat denial-of-service attacks in the context of
email spam, or server selection~\cite{Feldman2004short} by which pollers
only invite voters they believe will accept, could significantly reduce
the friction caused by this attack.  However, 
we have yet to explore whether
these defenses would be compatible with our goal of also protecting against
subversion attacks that operate by biasing the opinion poll sample toward
corrupted peers~\cite{Maniatis2003lockssSOSPshort}.

\subsection{Targeting the Effort Verification Filters}
\label{sec:effortful:poll}

To attack filters downstream of admission control, the
adversary must get through admission control as fast as allowable.  We consider
an attack by a ``brute force'' adversary
who continuously sends enough
poll invitations with valid introductory efforts to
get past the random drops; such invitations cannot arrive from credit or
even identities at the steady attack state, because they are more frequent
than what is considered
legitimate.  Since unknown peers suffer more 
random drops than peers in debt,
the adversary
launches attacks from in-debt addresses.  We conservatively
initialize all adversary addresses with a debt grade 
at all loyal peers.  We also give
the adversary an oracle that allows him to inspect all the loyal
peers' schedules.
This spares him the generation of introductory efforts that
would be wasted because of scheduling conflicts at loyal peers.

Once through admission control, the adversary can defect at
any stage of the protocol exchange: after providing
the introductory effort in the \msg{Poll} message (INTRO) by never
following up with a \msg{PollProof}, after providing the remaining
effort in the \msg{PollProof} message (REMAINING) by never following up
with an \msg{EvaluationReceipt}, and not defecting at all (NONE).

\begin{table}
\begin{tabular}{l||l|l|l|l}
Defection  & Coeff.     & Cost    & Delay   & Access \\
           & friction   & ratio   & ratio   & failure\\\hline\hline
INTRO      & $1.40$     & $1.93$  & $1.11$  & $4.99 \times 10^{-4}$\\
           & $1.31$     & $2.04$  & $1.10$  & $6.35 \times 10^{-4}$\\\hline
REMAIN-    & $2.61$     & $1.55$  & $1.11$  & $5.90 \times 10^{-4}$\\
ING        & $2.50$     & $1.60$  & $1.10$  & $6.16 \times 10^{-4}$\\\hline
NONE       & $2.60$     & $1.02$  & $1.11$  & $5.58 \times 10^{-4}$\\
           & $2.49$     & $1.06$  & $1.10$  & $6.19 \times 10^{-4}$\\
\end{tabular}
\caption{\label{tab:bruteforce}The effect of the brute force adversary
defecting at various points in the protocol on the coefficient of
friction, the cost ratio, the delay ratio, and the access failure
probability.  For each point, the upper numbers correspond to the 50-AU
collection and the lower numbers correspond to the 600-AU collection.}
\end{table}

Table~\ref{tab:bruteforce} shows that the brute
force adversary's most cost-effective strategy (i.e., with the lowest cost ratio
metric) is to participate fully
in the protocol; by doing so he is able to raise loyal peers'
preservation cost (i.e., their coefficient of friction) by a factor of $2.60$
($2.49$ for the large collection).
Doing so raises the baseline probability of
access failure to $6.19 \times 10^{-4}$ at a cost almost identical to
that incurred to the defenders (by a factor of $1.06$).
Fortunately, this continuous
attack even from a brute force adversary unconcerned by his own effort
expenditure is unable to
increase the access failure probability of the victims greatly; the
rate limits prevent him from bringing his advantage in resources to
bear.  Similar behavior in our earlier work~\cite{Maniatis2003lockssSOSPshort}
prevented a different unconstrained adversary from modifying the content
without detection.

In the analysis above, we conservatively assume that the brute force
adversary uses
attacking identities in the debt grade of their
victims.
Space constraints lead us to omit experiments with an adversary
whose minions may be in either even or credit grade.
This adversary polls a victim only after he has supplied that
victim with a vote,
then defects in any of the ways described above.
He then recovers
his grade at the victim by supplying an appropriate number of valid
votes in succession.
Each vote he supplies is used to introduce new minions that thereby
bypass the victim's admission control before defecting.
This attack requires the victim to invite minions into polls
and is thus rate-limited enough that it is less effective than brute
force. It is also further limited by the decay of first-hand reputation
toward the debt grade. 
We leave
the details for an extended version of this paper.

\section{Related Work}
\label{sec:relatedwork}

The protocol described here is derived from earlier
work~\cite{Maniatis2003lockssSOSPshort} in which we covered
the background of the LOCKSS system.
That protocol used redundancy, rate limitation, effort balancing, bimodal
behavior (polls must be won or lost by a landslide) and friend bias (soliciting
some percentage of votes from peers on the friends list)
to prevent powerful adversaries from
modifying the content without detection, or discrediting the
intrusion detection system with false alarms.  To mitigate its
vulnerability to attrition, in this work
we reinforce these defenses using admission control, desynchronization,
and redundancy, and restructure votes to support
a block-based repair mechanism that penalizes free-riding.
In this section we list work that
describes the nature and types of denial of service attacks,
as well as related work
that applies defenses similar to ours.

Our attrition adversary draws on a wide range of work in
detecting~\cite{Hussain2003short},
measuring~\cite{Moore2001short},
and combating~\cite{Argyraki2003short,Mahajan2001short,Savage2000short,Snoeren2001short}
network-level DDoS attacks capable of
stopping traffic to and from our peers.
This work observes that current
attacks are not simultaneously of high intensity, long duration,
and high coverage (many peers)~\cite{Moore2001short}.

Redundancy is a key to survival during some DoS attacks,
because pipe stoppage appears
to other peers as a failed peer.  Many systems use redundancy to mask
storage failure~\cite{Kubiatowicz2000short}. 
Byzantine Fault Tolerance~\cite{Castro1999short}
is related to the LOCKSS opinion polling mechanism in its goal
of managing replicas in the face of attack. It provides stronger
guarantees but is not as
well adapted to large numbers of replicas.
Routing along multiple redundant paths in Distributed Hash Tables (DHTs)
has been suggested as a way of increasing the probability that
a message arrives at its intended recipient despite nodes dropping
messages due to malice~\cite{Castro2002short} or pipe
stoppage~\cite{Keromytis2002short}.

Rate limits are effective in slowing the
spread of viruses~\cite{Forrest2000short,Williamson2002short}.
They have also been suggested for limiting the rate at which peers can
join a DHT~\cite{Castro2002short,Wallach2002short} as a
defense against attempts to control part of the hash space.
Our work suggests that DHTs will need to rate limit not only joins
but also stores to defend against attrition attacks.
Another
study~\cite{Saroiu2002OSDIshort} suggests that the increased
latency this will cause will not affect users' behavior.

Effort balancing is used as a defense against spam, which
may be considered an application-level DoS attack and has received
the bulk of the attention in this area.  Our effort balancing defense draws
on pricing via processing concepts~\cite{Dwork1992short}.  We measure cost by
memory cycles~\cite{Abadi2003short,Dwork2003short}; others use CPU
cycles~\cite{Back2002short,Dwork1992short}
or even Turing tests~\cite{SpamArrest}.
Crosby et al.~\cite{Crosby2003short} show that
worst-case behavior of application algorithms can be exploited in
application-level DoS attacks; our use of nonces and the bounded verification
time of MBF avoid this risk.  In the LOCKSS system we avoid strong peer
identities and infrastructure changes, and therefore rule out many
techniques for
excluding malign peers such as Secure Overlay
Services~\cite{Keromytis2002short}.



Related to first-hand reputation is the use of game-theoretic analysis
of peer behavior by
Feldman et al.~\cite{Feldman2004short}
to show that a reciprocative strategy
in admission control policy can motivate
cooperation among selfish peers.  

Admission control has been used to improve the usability of overloaded
services.  For example,
Cherkasova et al.~\cite{Cherkasova2002short}
propose admission control strategies that help protect long-running
Web service sessions (i.e., related sequences of requests) 
from abrupt termination.  Preserving the responsiveness of
Web services in the face of demand spikes is critical,
whereas LOCKSS peers need only manage their resources to make progress
at
the necessary rate in the long term. They can treat demand spikes
as hostile behavior.
In a P2P context, Daswani et al.~\cite{Daswani2002short}
use admission control (and rate limiting) to mitigate the effects
of a query flood attack against superpeers in unstructured
file-sharing peer-to-peer networks such as Gnutella.

Golle and Mironov~\cite{Golle2001short} provide compliance enforcement in
the context of distributed computation using a receipt technique similar
to ours.  Random auditing using challenges and hashing has been
proposed~\cite{Cox2003short,Wallach2002short} as a means of enforcing trading
requirements in some distributed storage systems.

In DHTs waves of synchronized routing updates caused
by joins or departures cause instability during periods of high
churn.  Bamboo's~\cite{Rhea2004short} desynchronization defense
using lazy updates is effective.





\section{Future Work}
\label{sec:futurework}

We are currently exploring the admission control parameter space.  In
particular, we are studying the effects of varying the length of the
refractory period, the drop probabilities for unknown and in-debt
peers, and the effects of running the audit protocol over a larger
number of AUs.  As the number of AUs increases and peers are naturally
more busy participating in polls, a longer refractory period may be
more appropriate to allow loyal peers time to handle the load of polls
called by other loyal peers.

We have three immediate goals for future work.  
First, we observe that although the protocol is symmetric, the
attrition adversary's use of it is asymmetric.  It may be that
adaptive behavior of the loyal peers can exploit this asymmetry.  For
example, loyal peers could modulate the probability of acceptance of a
poll request according to their recent busyness.  The effect would be
to raise the marginal effort required to increase the loyal peer's
busyness as the attack effort increases.  Second, we need to
understand how our defenses against attrition work in a more dynamic
environment, where new loyal peers continually join the system over
time. Third, we need to consider combined adversary strategies; it could
be that the adversary can use an attrition attack to weaken the system
in some way that leaves it more vulnerable to other attack goals.

\section{Conclusion}
\label{sec:conclusion}

The defenses of this paper equip the LOCKSS system to resist attrition well:
\begin{itemize}
\item Application-level attrition attacks, even from adversaries with no
resource constraints and sustained for two years, can be defeated with
a constant factor of over-provisioning. Such over-provisioning is natural
in our application, but further work may significantly reduce the required
amount.

\item The strategy that provides an unconstrained adversary with the greatest
impact on the system is to behave as a large number of new loyal peers.

\item Network-level attacks do not affect the system significantly
unless they are (a) intense enough to stop all communication between
peers, (b) widespread enough to target all of the peers,
and (c) sustained over a significant fraction of an inter-poll interval.
\end{itemize}

Digital preservation is an unusual application, in that the goal is to
prevent things from happening.  The LOCKSS system resists failures and attacks from powerful
adversaries \emph{without} normal defenses such as long-term secrets and
central administration.   The techniques that we have developed
 may be primarily applicable to preservation, but 
we hope that our conservative design will assist others in building systems that
better meet society's need for more reliable and defensible systems.
 
Both the LOCKSS project and the Narses simulator are hosted at
SourceForge, and both carry BSD-style Open Source licenses.
Implementation of this protocol in the production LOCKSS system is in
progress.

\section{Acknowledgments}
\label{sec:acknowledgments}
We are grateful to Kevin Lai, Joe Hellerstein, Yanto Muliadi,
Geoff Goodell, Ed Swierk, and Lucy Cherkasova for their help.
We are especially thankful to Vicky Reich, the director of the LOCKSS
program.

This work is supported by the National Science Foundation (Grant No.\
9907296) and by the Andrew W. Mellon Foundation.
Any opinions, findings, and conclusions or
recommendations expressed here are those of the authors and do not
necessarily reflect the views of these funding agencies.

{\footnotesize \bibliographystyle{acm} \bibliography{../common/bibliography}}

\end{document}